# Port Forwarding Services Are Forwarding Security Risks

Haoyuan Wang[1], Yue Xue[1], Xuan Feng[2], Chao Zhou[1], and Xianghang Mi[§1]

[1]University of Science and Technology of China
[2]Microsoft Research Asia
https://chasesecurity.github.io/Port-Forwarding-Services

*Abstract*—We conduct the first comprehensive security study on representative port forwarding services (PFS), which emerge in recent years and make the web services deployed in internal networks available on the Internet along with better usability but less complexity compared to traditional techniques (e.g., NAT traversal techniques). Our study is made possible through a set of novel methodologies, which are designed to uncover the technical mechanisms of PFS, experiment attack scenarios for PFS protocols, automatically discover and snapshot port-forwarded websites (PFWs) at scale, and classify PFWs into well-observed categories. Leveraging these methodologies, we have observed the widespread adoption of PFS with millions of PFWs distributed across tens of thousands of ISPs worldwide. Furthermore, 32.31% PFWs have been classified into website categories that serve access to critical data or infrastructure, such as, web consoles for industrial control systems, IoT controllers, code repositories, and office automation systems. And 18.57% PFWs didn't enforce any access control for external visitors. Also identified are two types of attacks inherent in the protocols of Oray (one well-adopted PFS provider), and the notable abuse of PFSes by malicious actors in activities such as malware distribution, botnet operation and phishing.

## 1. Introduction

The increasing scarcity of IPv4 addresses leads to a tension between offering web services to the public and the costly cloud computing services. To address this tension, various techniques have been explored so as to *port forward* external visits to web services hosted in networks that are either private or have only dynamic public IP addresses. Two typical examples of such solutions are NAT traversal techniques [1] and dynamic DNS services [2]. NAT traversal techniques aim to punch a hole in the gateway NAT device so that network connections towards an internal server can be established. In addition to being heterogeneous and complicated, NAT traversal techniques either require privileged permissions to proactively inject entries in the NAT table, or rely on external servers (e.g., a STUN server or a TURN server) to establish or maintain the connection.

§. Corresponding author at xmi@ustc.edu.cn

Also, their effectiveness varies a lot across NAT devices and transport layer protocols, which is further undermined by the increasing adoption of symmetric NATs [1]. Furthermore, for networks with dynamic public IP addresses, dynamic DNS services have been adopted to timely update the DNS records when the public IP address has changed. However, the latency incurred by detecting IP change and propagating the DNS updates can still undermine the availability of the internal web services.

As an alternative solution to address these limitations, port forwarding services (PFS) emerge in recent years, which avoid the complexity of NAT traversal techniques, get rid of the need for dynamic DNS services, and provide full-fledged and user-friendly port forwarding for any internal network service. Representative PFS providers include Ngrok[3], Oray[4], and Portmap.io[5]. At a high level, to port-forward an internal website (PFW), a PFS agent program will be installed in the internal network (e.g., on the same server with the PFW) to proactively initiate one or more persistent connections (PFW tunnels) with the external PFS server. Also, each PFW will be automatically assigned with a unique PFW domain name, which is resolved to the external PFS server. Once setup, visits towards the PFW domain name will be forwarded by the PFS server, via the PFS agent, to the internal website, and vice versa.

However, given the emergence and increasing adoption of PFS [6], [7], little is known about how PFS works from a technical perspective and to what extent the underlying PFS protocols are vulnerable to network attacks. It is also unclear regarding what kinds of websites have been port-forwarded by PFS, whether such a forwarding is well authorized by the PFW administrator, and to what extent a PFW is well protected against unsolicited visitors. Also, considering several abuse incidents from miscreants [8], [9], it is interesting to comprehensively profile the abuse of PFS and its respective security implications. In this paper, we report the first of its kind security study on PFS, which has answered these research questions with robust measurement methodologies, interesting findings and observations, novel attacks, as well as promising mitigation proposals.

In our study, we target two representative PFSes, namely, Ngrok and Oray, both of which are among the most adopted PFSes along with rich port forwarding features. And our study is made possible through a set of novel methodologies.

First of all, a PFS testbed has been built up, to understand how a PFS works from a technical perspective, and experiment potential attacks. In the design of this testbed, several websites hosted in an internal network are port-forwarded to the public via both Oray and Ngrok. And the testbed is equipped with a set of analysis modules for traffic capturing and attack experiments. In addition, to automatically discover and snapshot PFWs, a PFW collector has been designed wherein randomly generated PFW domain names are discovered by querying passive DNS, and headless browsers are instrumented to capture the snapshots of PFWs in an efficient and distributed manner. Given the millions of PFW snapshots captured through this collector, it is challenging to profile their categories. To conquer this challenge, a multiclass machine learning classifier is further developed, which takes multi-modal elements (textual content and visual elements) of a PFW snapshot as the input, and outputs the pre-defined website category it belongs to.

Leveraging these methodologies, our study has distilled a set of novel findings and observations, which are summarized as below.

First of all, PFSes have been adopted to port-forward millions of internal websites distributed globally. Leveraging the PFW collector, we carried out PFW collection across more than 5 months between June 2022 and December 2022, through which, 6,865,169 PFW domain names have been observed, and 275,513 were found to be reachable and had snapshots captured. In total, we have captured 3,501,556 PFW snapshots. Also, among PFWs of Ngrok, 261,765 have the public IP addresses of their internal networks identified (§4.2), which includes 63,264 unique IPv4 addresses and 20,234 unique IPv6 addresses, which suggests these PFWs are widely distributed across 173 countries and 5,414 ISPs.

Furthermore, 32.31% PFWs turn out to be web interfaces that are used to either access critical data or control critical infrastructure, and forwarding such kinds of PFWs to the public may incur non-negligible security risks. Such PFWs include industrial control systems (1.83%), web consoles for IoT controllers and devices (1.11%), web consoles for network devices (3.48%), office automation systems (14.56%), data stores (2.46%), and even source code repositories (0.79%). What is even more worrying is that, among such PFWs, 18.57% failed to enforce any authentication, 77.85% enforced only username/password authentication, and only 8.00% required challenge-response tests for login attempts.

Besides, we have identified two types of attacks in Oray's PFS protocols, which have been demonstrated through attack experiments on our PFS testbed. One attack exists in the data plane protocol which allows attackers on any intermediate hop between the Oray agent and the Oray server to perform a man-in-the-middle (MITM) attack and manipulate requests towards and responses from a given PFW. Another attack targets the control plane communication, which allows an attacker to use the Oray agent program as a stepping stone to attack internal web infrastructure co-located with the PFW. We have responsibly disclosed these two vulnerabilities to Oray, which in turn has well acknowledged both vulnerabil-ities along with a bug bounty rewarded to us.

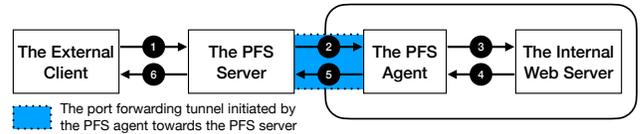

Figure 1. The PFS usage scenario.

Also, port forwarding services are being abused to a concerning extent and in various malicious activities, particularly, malware distribution, phishing & fraud, and tunneling communications between compromised machines and the command & control servers of remote access trojans (e.g., njRAT). As the result, 76% Ngrok PFW IPs have been detected as malicious by one or more detection engines while over 55% have been further detected as malicious by ten or more detection engines. Also, such extensive abuse of PFS also incurs a non-negligible challenge for existing defense systems, particularly considering that malicious PFWs can emerge and migrate away very quickly while the underlying attacking servers are kept hidden.

Our contributions can be outlined as follows.
• We conduct the first extensive security study on representative port forwarding services.
• A novel methodology has been proposed and implemented to automatically discover, snapshot, and classify port-forwarded websites.
• A set of novel security findings on port-forwarded websites (PFWs) have been distilled along with supportive analysis and data points.
• Two attack scenarios have been identified and demonstrated for the port forwarding protocols of Oray, one of the most adopted PFS providers.

## 2. Background

**Port forwarding services (PFS)**. A PFS is designed as an out-of-the-box service to help ordinary users carry out port forwarding tasks. It circumvents the necessity of configuring the gateway NAT device by reverse port forwarding, i.e., a tunnelling technique which allows you to forward traffic from the outside world to your local machine (e.g., reverse SSH tunneling). In the scenario of a PFS, a program, namely, the PFS agent, is deployed to be co-located inside the same private network with the internal network service that needs to be forwarded, and is instructed to proactively set up one or more persistent tunnels to a PFS server. Given the port forwarding tunnels are set up, the traffic visiting the PFS server will be redirected to the PFS agent. Figure 1 depicts how the client-side web traffic (e.g., an HTTP request) traverses the PFS server, PFS agent, before reaching the internal network service, which tends to be a website (PFW). To distinguish among PFWs, the PFS provider assigns a unique domain name to each PFW, which tends to be a randomly generated subdomain of an apex domain registered under the PFS provider.

TABLE 1. The list of port forwarding services.

| Provider Name | Service Website | PFWs [1] | pDNS I [2] | pDNS II [3] |
|---|---|---|---|---|
| Ngrok | *ngrok.com* | 3.47M | 125M | 87M |
| Oray | *hsk.oray.com* | 4.60M | 43M | 104K |
| Portmap.io | *portmap.io* | 3K | 8M | 8M |
| NATAPP.cn | *natapp.cn* | 225K | 46M | 51K |
| localhost.run | *localhost.run* | 4K | 594K | 2M |

[1] PFWs as observed during 2022 in passive DNS (pDNS).
[2] The historical DNS queries as observed by QiAnXin pDNS by July 2023.
[3] The historical DNS queries as observed by RISKIQ pDNS by July 2023.

Our study has identified 5 PFSes in total and the full list can be found in Table 1. We have further investigated these services and compared them from various aspects, particularly, the port forwarding features and the scale in terms of the number of PFWs. As detailed in §3.2, a PFW domain name is usually a unique subdomain of one or more apex domains registered under the PFS provider. To estimate the scale of each PFS provider, we manually identified its apex domains for addressing PFWs, and queried two passive DNS services to collect all the subdomains (i.e., *gicp.net* for Oray) that were active during 2022. As shown in Table 1, Oray and Ngrok have the largest number of PFWs observed during 2022, 4.6M/3.5M respectively. Oray and Ngrok are also among the most visited service websites in terms of historical DNS queries of the service domain name (e.g., *ngrok.com* for Ngrok). Besides, Ngrok is featured by not only its high popularity and also global reach, while Oray is dedicated to customers in China and claims to have been adopted in many areas (e.g., video surveillance, remote device management) and by many large IT vendors [10], e.g., Cisco, Huawei, SIEMENS, TP-Link, and Hikvision. We therefore chose Ngrok and Oray as the targets of our study. And we believe that many of our measurement results on Oray and Ngrok are applicable to other services, e.g., the forwarding protocols, the categories of PFWs, and the potential exposure of sensitive internal web services, etc.

**Passive DNS**. Passive DNS (pDNS) datasets store historical DNS queries and responses that are collected from widely distributed DNS resolvers. A typical pDNS record consists of a DNS record, the timestamps when it was first and last observed, as well as the number of DNS queries towards this record as aggregated from DNS resolvers. In our study, we have utilized two passive DNS datasets, with one provided by QiAnXin[1] and the other sourcing from RISKIQ[2]. Combining both datasets was found to lead to a higher coverage for PFW domain names.

## 3. Methodology

Upon the background knowledge, we present in this section the research methodology which consists of three modules: the PFS analyzer, the PFW collector, and the PFW classifier. As the first step of our study, we try to understand how PFS works and its potential security risks, which is made possible by a PFS testbed, as elaborated in §3.1. Given a set of security risks identified for PFS protocols, we are motivated to profile what internal websites have been forwarded to the public. Therefore, a PFW collector is designed and implemented to automatically discover newly emerged PFW domain names and capture snapshots for each PFW, as detailed in §3.2. Given a large scale of PFW snapshots collected, a PFW classifier is built up to automatically decide what category a PFW snapshot belongs to, which is presented in §3.3.

### 3.1. The PFS Testbed

To uncover how PFS works and test potential vulnerabilities, a PFS testbed was built up. In the design of this testbed, multiple websites deployed in internal networks are port-forwarded to the public via both Oray and Ngrok. Along the websites and PFS agents, tcpdump [11] is enabled to capture network traffic involving the PFS agent, in an attempt to uncover what network protocols are utilized to fulfill the PFS communication, e.g., how the PFS agent communicates with the PFS servers and the internal websites. Besides, the mitmproxy [12] tool is deployed to evaluate potential risks of man-in-the-middle (MITM) attacks.

Leveraging this testbed, we triggered diverse traffic towards PFWs under our control from various locations. Also, we modified configurations available on the PFS web console to observe how such configuration updates could be synced to the PFS agents. Besides, when evaluating potential MITM attacks, we redirected the PFS agent traffic to the mitmproxy by configuring the firewall settings of the host where the PFS agent was installed. Note that when analyzing PFS vulnerabilities through MITM, the traffic flows under interception was generated by our own browser and tunnelled to internal websites under our control. We believe there are no ethical issues as our experiments have no impact on any third parties. Empowered by this testbed, we have successfully uncovered many technical details of the PFS ecosystem as well as demonstrated multiple security vulnerabilities. For more details, see §4.1 for the technical details and §5.2 for the vulnerabilities.

### 3.2. The PFW Collector

To get a deep understanding of the PFS ecosystem, it is critical to know what websites have been tunnelled by PFS, especially considering that port forwarding a website is the most common use case of PFS. We refer to such websites as port-forwarded websites (PFWs). Unlike a typical website, a PFW is more dynamic and can appear and disappear very quickly. In addition, the domain name of the PFW is a randomly generated subdomain under the apex domain names of PFS, and no solutions exist to enumerate all publicly available PFWs. To address these issues, we design a PFW collector to automatically discover PFW domain names through querying passive DNS, and snapshot the

1. http://en.qianxin.com/
2. https://www.riskiq.com/

discovered PFWs in an efficient and distributed manner. Next, we introduce more details about this PFW collector.

**Discovering PFW domain names**. The first step of our PFW collector is to discover PFW domain names in a timely manner. As learned from PFS documents and our empirical experiments, a PFW domain name is usually a unique subdomain of one or more apex domains registered under the PFS provider. For instance, a PFW of Ngrok can be uniquely identified by a subdomain of *ngrok.io*, while *random-string.oray.com.cn* is a PFW domain pattern of Oray where *oray.com.cn* is an apex domain registered under Oray and *random-string* is a sequence of randomly generated characters to uniquely identify a given PFW. Also, a PFS provider can address its PFWs under many different apex domains (e.g., *vicp.net* is another apex domain of Oray used to address PFWs). Therefore, to discover domain names of PFWs, we first need to identify all such kinds of PFS apex domains, which are then queried against passive DNS to learn their subdomains (i.e., PFW domain names).

However, it is non-trivial to identify all PFS apex domains, since neither Ngrok nor Oray provides the full list of PFS apex domains they have deployed. Our observation is that both Ngrok and Oray use a small set of IP addresses to exclusively host their apex domains. Therefore, we adopted a snowballing strategy to automatically reverse all apex domains behind these IP addresses owned by each PFS provider. Specifically, we use a small set of apex domains confirmed manually as seed to query passive DNS records to identify their hosting IP addresses. Given these newly identified hosting IP addresses, we reversely looked up the passive DNS records to identify any apex domains hosted on these IPs. This process continued until no more apex domains or IP addresses could be identified. In total, we identified 82 unique apex domains for Oray and only one apex domain (ngrok.io) for Ngrok. We double-checked the case of Ngrok through service trials and manual passive DNS lookup, and have thus confirmed that it indeed hosts all PFWs under the single apex domain.

Given these apex domains, we set up a daily job to query pDNS datasets and extract subdomains of these apex domains. Our PFW collector focuses on profiling *active* PFWs, most if not all of which should have active DNS queries on a weekly base. Therefore, subdomains will be considered inactive and excluded from our collection if their passive DNS records are not active within the last 7 days at the time of the query. By then, the left ones are considered as active PFW domain names and will be further visited to capture their snapshots.

**Testing aliveness of PFW domain names**. Given PFW domain names with DNS queries observed in the last 7 days, the next step should be to dynamically visit their webpages and capture their content as a snapshot. However, dynamically loading a webpage can be costly in terms of computing and bandwidth. Thus, we designed a prior step to quickly filter out PFWs that are inactive. This is achieved through sending HTTP GET requests over both HTTP and HTTPS protocols. For each test request, the connection timeout is configured as 10 seconds, and a PFW website is considered alive as long as it is reachable with HTTP(S) response received, regardless of the HTTP response code. Note that a PFW with non-200 HTTP response is also deemed active. Through this aliveness check, we can significantly lower the workload of our next step (PFW snapshotting) by 96%.

**Snapshotting PFWs**. Given PFWs that are tested to be alive (or visitable), our collector dynamically visits each PFW in a headless browser to record the network traces, render the landing page, and capture a screenshot for the rendered webpage, which together are considered as a snapshot for a PFW. Note that our collector only visited the landing page (the homepage) for each PFW due to several factors. On one hand, for most PFWs, the landing page contains enough information to decide which security-sensitive category they belong to. On the other hand, many PFWs are used to control IoT or cyber-physical devices, and visiting subpages of such PFWs can potentially lead to control commands being sent to the respective device. Similarly, many PFWs were found to be data stores or code repositories, visiting their subpages may expose very sensitive data, but distill little or no research value.

Our PFW collector is implemented in Python using several libraries, in particular Requests for programmable HTTP requests, and Playwright [3] for dynamic and automatic website rendering. In terms of deployment, our collector was set up to run on a daily basis, as PFWs can emerge and disappear very quickly. To profile the availability of PFWs across countries, the same collector was distributed across three countries: China, the USA and Germany. We ran our collector between June 2022 and December 2022. In total, we have captured 3,501,556 snapshots for 275,513 unique PFWs.

### 3.3. The PFW Classifier

Once PFWs have been captured, we move on to profile their categories, in an attempt to understand what kinds of websites have been exposed as PFWs. To handle the large volume of over 3.5 million PFW snapshots, a multiclass classifier is designed to take a PFW snapshot as the input and predict it as one of a set of commonly observed categories. One thing to note, rather than serving as a generic website classifier, this classifier is tailored for PFW categories that have been frequently observed during our manual study of PFW snapshots, e.g., industrial control systems (ICS) and office automation systems (OA). Below, we give more details about this classifier.

**Groundtruth**. To create the groundtruth, two members of the research team independently labeled a common set of PFW snapshots. During the labeling process, the two labelers periodically synced with each other to resolve any labeling conflicts. Initially, we manually labeled 1,941 samples of 12 categories. As the sample volume for some categories is still too small to train a robust classifier, we moved forward to train a weak classifier and used this weak classifier to predict and validate unlabeled PFS snapshots in order to enlarge and balance our groundtruth dataset. Specifically, to enlarge samples for a given category, a PFW snapshot would be selected out for manual validation if its prediction probability is higher than 0.2 for the given category. Through this process, our final groundtruth dataset is expanded to 5,513 samples of 12 categories, as listed in Table 2.

---

3. https://playwright.dev/

TABLE 2. THE DISTRIBUTION OF PFWS IN THE GROUNDTRUTH ACROSS THEIR CATEGORIES.

| Category | % PFWs | Category | % PFWs |
| --- | --- | --- | --- |
| ICS [1] | 7.04% | IoT [2] | 5.66% |
| Network Devices | 10.19% | Remote Desktop | 11.54% |
| Office Automation (OA) | 9.61% | Data Store | 11.01% |
| Code Repository | 5.30% | NAS [3] | 7.78% |
| Webserver Default Page | 5.95% | Error Page | 5.64% |
| Blank Page | 7.78% | Others | 12.50% |

[1] ICS is short for Industrial Control Systems.
[2] IoT denotes web consoles for IoT Controllers and Devices.
[3] NAS is short for network-attached storage.

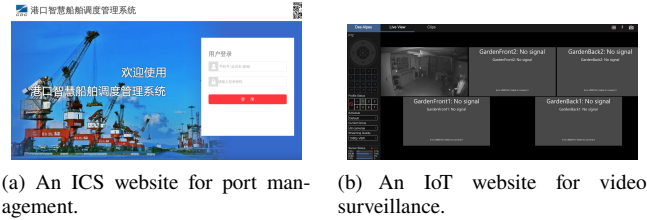

(a) An ICS website for port management.

(b) An IoT website for video surveillance.

Figure 2. PFW cases illustrating the importance of visual elements.

Among these categories, some deserve more explanation. One is ICS versus IoT. ICS is defined as web interfaces to control or monitor industrial control systems (i.e., cyber-physical systems), such as the SCADA (supervisory control and data acquisition) system in the oil/gas plant and pipeline automation, the PLCs (programmable logic controllers) in transportation systems, or remote terminal units (RTUs) in remote power plants and substations. Then, a PFW is considered as IoT when it serves as a web console to monitor or control devices that are deployed in smart homes or smart buildings. However, some PFW cases sit on the borderline between ICS and IoT, for which we take a conservative strategy and tend to label such cases as IoT.

**Preprocessing PFW snapshots.** As detailed above, a PFW snapshot consists of a screenshot of the landing page and DOM objects (e.g., Javascript files, images, etc) loaded when rendering the landing page. Given a PFW snapshot, some preprocessing steps are applied to extract all the text elements embedded in the screenshot and DOM objects. Firstly, visual text elements are extracted from the screenshot leveraging Tesseract [13], an open-source tool for multilingual optical character recognition (OCR). Then, these visually important text elements are complemented by those identified in the DOM objects, such as the HTML file and JavaScript snippets. We consider both because the text elements in the screenshot are visually important while those in DOM objects may contain hidden information such as the organization name and webpage description. Besides the modality of the text elements, our classifier also considers the screenshot as the visual modality.

**Training and evaluating the multi-modal PFW classifier.** There are two critical issues to consider when building up the classifier. As illustrated in Figure 2(a) and Figure 2(b), both the text and the visual elements play an important role in terms of deciding the website category. Therefore, our classifier needs to support both visual and text modalities. In addition, for the text modality, PFWs consist of multiple natural languages, and the classification model should support multilingual inputs. To meet these requirements, our PFW classifier is built up through adopting and fine-tuning LayoutXLM [14], a pre-trained multi-modal model which has achieved SOTA results for many multilingual document understanding tasks. Inside the architecture of LayoutXLM, 12 layers of transformer blocks are used with 12 self-attention heads and a hidden size of 768, which sums up to a total number of 345M parameters. During training, we explored various combinations of hyperparameters and the final combination with the best performance is a batch size of 4, a gradient accumulation step of 16, and an epoch number of 10.

As a result, our model has achieved 95% for the micro average accuracy/recall/precision. Also, when considering the top 3 predictions for each sample, the recall can be further improved to 98%. The category-wise performance results are also listed in Table 3. As we can see, the performances for most categories are high enough. We then further looked into the false classifications and found that some OA PFWs were predicted as ICS, probably because they belong to organizations that operate industrial control systems, e.g., the information system of a construction company, a company providing maintenance services for the power grid, etc. The ICS-relevant text or visual elements in such OA PFWs may have confused the classifier. Also, regarding false positives for the OA category, we found that most belong to either data store or ICS and some can indeed be assigned with multiple labels, e.g., a PFW used to manage employees and contracts can be considered as both a data store and an OA system. We leave it as our future work to explore multi-label PFW classification.

### 3.4. Ethical Considerations

We take ethics seriously and have carefully designed our methods to avoid any ethical issues. Specifically, we first attempted to apply for IRB review, but later found that this is not feasible due to the limited scope of our institution's IRB, which primarily focuses on biological/medical studies and has yet to be capable of reviewing applications from other domains, including cybersecurity. Instead, we consulted several cybersecurity researchers and adopted the best ethical practices established by previous studies [15], [9], [16]. As the result, when capturing snapshots for PFWS, our collector only visits the landing page of each PFW rather than any subpages. We design the collector in this manner according to the observation that some PFWs are consoles for industrial control systems or IoT devices without access control, and visiting links embedded on the landing page may lead to operations on a physical device, but the landing page always works as the entrance or the menu which involves little sensitive operation. Similarly, when manually visiting a PFW during case studies and PFW labeling, we were very cautious and avoided any unnecessary or unpredictable subpage visits. Besides, all the PFS snapshots were stored securely on a research server with strict access control. We have also made responsible disclosures to relevant parties including PFW administrators and PFS providers, in light of security findings, as detailed in §7.

TABLE 3. Class-wise performance of the PFW classifier.

| Category | Precision | Recall | F1-Score |
|---|---|---|---|
| Industrial Control System (ICS) | 95% | 97% | 96% |
| IoT Controller and Devices | 100% | 90% | 95% |
| Network Devices | 97% | 97% | 97% |
| Remote Desktop | 99% | 99% | 99% |
| Office Automation (OA) | 90% | 90% | 90% |
| Data Store | 98% | 95% | 96% |
| Code Repository | 97% | 98% | 97% |
| Network-attached Storage (NAS) | 99% | 96% | 97% |
| Webserver Default Page | 100% | 94% | 97% |
| Error Page | 93% | 98% | 96% |
| Blank Page | 99% | 100% | 100% |
| Others | 80% | 85% | 82% |

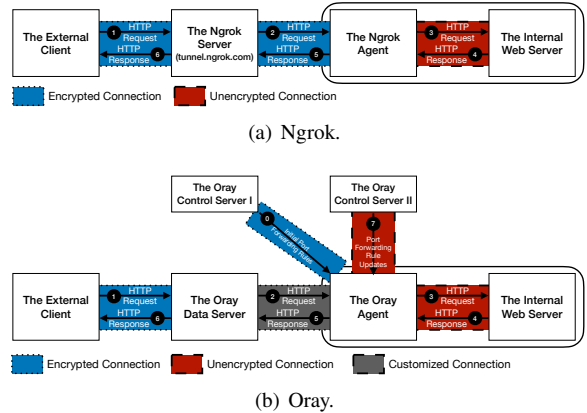

(a) Ngrok.

(b) Oray.

Figure 3. The port forwarding protocols.

### 3.5. Limitations

Several limitations exist in our methodology. First of all, our study focuses on how a PFS port-forwards *web services or websites* to the public, and our findings on PFWs may not be applicable to other port-forwarded network services, e.g., SSH servers. Besides, our PFW collector can only capture a PFW if its domain name has been observed by the passive DNS datasets. Therefore, the coverage of our collector for PFWs is constrained by the two passive DNS datasets adopted in our study. Also, as passive DNS datasets are known to bias towards websites of a longer lifespan or higher access frequency, our PFW collector inherits this limitation, i.e., long-lived and frequently accessed PFWs are more likely to be captured by our collector. However, our collector is designed in a manner that the passive DNS datasets can be seamlessly replaced or complemented by any other domain datasets, e.g., domains recorded in certificate transparency, so as to mitigate this limitation. Furthermore, A PFW can emerge and disappear quickly, which means our daily deployed collector could still miss such short-lived PFWs even if their domain names got recorded by the passive DNS datasets. Furthermore, many PFWs involved in malicious activities (e.g., spam and phishing) may migrate very fast (less than one day), making our daily collection jobs incapable of capturing valid snapshots for them. Fortunately, by integrating the reports obtained from threat intelligence platforms, we are still able to effectively profile them (see §6).

## 4. PFS Providers and Port-Forwarded Websites

In this section, we reveal for the first time the technical mechanisms of two representative PFS providers, together with a comprehensive measurement for PFS websites (i.e., PFWs). The distilled knowledge serves as a basis for reasoning about and evaluating the potential security risks.

### 4.1. PFS Providers

As detailed in §3.1, we selected two representative PFS providers, exposed self-deployed internal websites through their services to the public, and analyzed the captured network traffic. Through these steps, we are able to successfully uncover their technical mechanisms, customer vetting policies, and pricing policies, as detailed below.

**Technical mechanisms**. Regarding the technical mechanisms, we focus on how the PFS agent works from the network perspective, which is hidden from PFS customers and may vary across PFS providers. A PFS agent, the software program installed on an internal server, is configured to relay external visits to an internal web service. As shown in Figure 1, it sits between the PFS server and the internal web service, and intermediates traffic between them. For both Oray and Ngrok, the network traffic of the PFS agent can be grouped into two categories: the control plane communication and the data plane communication.

For Ngrok, as shown in Figure 3(a), once the agent program is started, it connects to the Ngrok server (tunnel.ngrok.com) through a single persistent HTTPS connection, which serves both as the data plane relaying traffic to and from the internal web service, and as the control plane transmitting control data (e.g., the authorization token) between the agent and the Ngrok server. When multiple PFWs are forwarded by the same Ngrok agent, they share the same HTTPS tunnel connection. Given the tunnel connection setup, when an HTTP request arrives at the agent, it will be further forwarded to the internal web server through a plaintext HTTP connection. Also, inside the HTTP request, two headers have been added by the Ngrok server so as to inform the internal web server about the visitor. Specifically, the header named as *X-Forwarded-For* is used to store the source IP address of the forwarded HTTP request, while the other header named *X-Forwarded-Proto* denotes the protocol through which the HTTP request is sent to the Ngrok server.

In Oray, a more complicated interaction process is implemented so as to support more flexible functionalities, such as dynamically updating the port forwarding rules without the need of stopping and restarting the Oray agent. Specifically, as illustrated in Figure 3(b), the Oray agent first queries the Oray control server (hsk-embed.oray.com) through an HTTPS connection, and extracts a set of configurations including which local service to forward traffic to, which Oray data server to connect to set up the tunnel connection, among others. Upon these configurations, the agent moves to setup a long-lived TCP connection with the specified Oray data server (e.g., phfw-overseasvip.oray.net at TCP port 6061) as the data tunnel, to relay HTTP traffic between the internal web service and the external visitor. The Oray agent will also periodically send heartbeat UDP packets to the Oray data server. Inside the data tunnel, the HTTP requests and responses are transmitted through a proprietary application-layer protocol. However, as revealed and demonstrated in §5, the relayed HTTP traffic is not well protected, giving attackers the opportunity to do a full-fledged man-in-the-middle (MITM) attack.

Oray also provides a web console for its customers to update the tunneling configurations, e.g., change which local service to forward the traffic to without interrupting the tunnel. To achieve this, the agent sets up another long-lived TCP connection with another Oray control server (e.g., phsle5-adv01.oray.net at port 6061) and uses it as the control plane for the server to push configuration updates back to the agent. Both control planes, the one to pull initial configurations and the other one to pull configuration updates, have non-negligible security vulnerabilities that allow the attacker to disrupt the port forwarding, e.g., to visit co-located internal services that are not intended to be exposed to the public, as detailed in §5.

**Access control for PFWs**. Both Ngrok and Oray allow paid users to enable IP-based access control. An user can configure an IP allowlist or blocklist so as to restrict which source IPs can (not) access the respective PFW. For IPs that are denied from access, Ngrok returns HTTP 403 with the error code ERR_NGROK_3205 while Oray drops the connection directly. Besides, Ngrok also provides HTTP Basic Auth and User Agent Filter for free users, and a denied visit will receive HTTP 401 with no error code and 403 with ERR_NGROK_3211 respectively. However, as observed from PFW snapshots, many PFWs adopt none of these access control measures, leading to non-negligible risks.

TABLE 4. THE STATISTICS OF PFW COLLECTION.

| PFS Provider | PFWs in pDNS | Reachable PFWs | Snapshots |
|---|---|---|---|
| Oray | 4,033,136 | 229,403 | 2,966,651 |
| Ngrok | 2,832,033 | 46,110 | 534,905 |
| All | 6,865,169 | 275,513 | 3,501,556 |

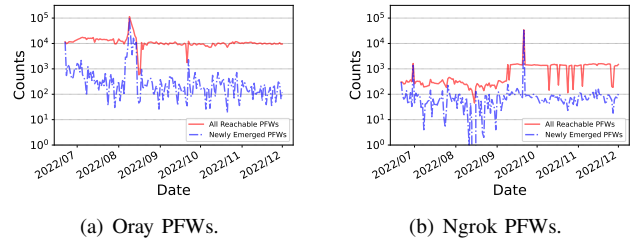

(a) Oray PFWs.   (b) Ngrok PFWs.

Figure 4. The daily fluctuation of reachable PFWs.

**Customer vetting policies**. As learned from our experience of using both services, Ngrok doesn't enforce any background check process for its new customers, anyone can easily register a Ngrok account and start tunneling traffic since then. However, Oray requires each customer to verify their identity by uploading photos of the ID card as well as passing live face recognition [4].

### 4.2. Port-Forwarded Websites

Upon the knowledge distilled on PFS providers, we then move on to profile the port-forwarded websites (PFWs) with a focus on their scale, evolution, and categories.

**Scale**. Below, unless otherwise noted, we measure PFWs that were observed between June 21, 2022 and December 1, 2022. In total, during the more-than-5-month collection period, 6,865,169 PFW domain names were observed in the passive DNS datasets, among which 275,513 were found to be reachable and had snapshots captured. As a result, we have captured 3,501,556 PFW snapshots. Table 4 also lists the statistics per PFS provider.

**Evolution**. Figure 4 presents how the daily collected PFWs fluctuate across the collection period. For both PFS providers, we can see that new PFWs are constantly appearing, while the daily reachable PFWs don't show a stable upward trend, which means that the churn rate for PFWs is high and that many PFWs may have a short lifespan. We can also see a sudden increase in Ngrok snapshots in September, due to the integration of a new passive DNS dataset from RISKIQ. Some data points look abnormally lower than others, which is due to server crashes. On average, we observe 18,426 PFWs per day for Oray, and 3,262 for Ngrok.

**Lifetime**. Given the scale and evolution of PFWs, we also profile how long a PFW keeps alive (online) since it is first observed. Here, we define two metrics to profile this question. The first is the lifetime $lt$ of a PFW as $lt =$ $date_{last} - date_{first}$, i.e., the interval in days between the first date and the last date that a PFW is observed. However, a PFW may be offline for some of the time between $date_{first}$ and $date_{last}$. We therefore define another metric, namely activeness time $at$, as $\sum_{d=first}^{last} active_d$, where $active_d$ is 1 if the PFW is active on date $d$, otherwise $active_d$ is 0.

---

4. https://www.oray.com/announcements/affiche/?aid=604

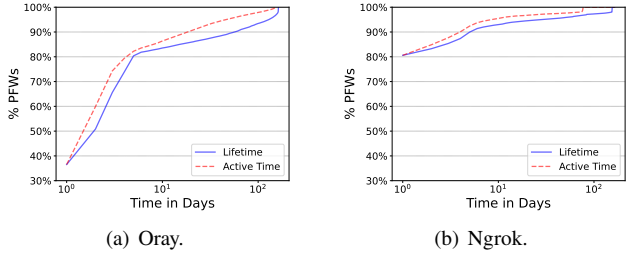

Figure 5. The cumulative distribution of PFW over their lifetime and activeness time.

The cumulative distribution of PFWs across their lifetime and activeness time are presented in Figure 5(a) for Oray and Figure 5(b) for Ngrok. Different from a typical online website, PFWs tend to have a short lifetime. Specifically, during our collection period of 164 days, 83.54% PFWs of Oray and 92.97 % PFWs of Ngrok have a lifetime shorter than 10 days. And 50.83% of Oray and 83.27% of Ngrok have a lifetime of less than 2 days. In addition, across two PFS providers, 29.09% of PFWs have an activeness time shorter than their lifetime and 12.69% PFWs have an activeness time shorter by 5 or more days. This implies that many PFWs are not always online during their lifetime due to various factors, e.g., when the device (e.g., a laptop) hosting the PFS agent gets offline.

**The adoption of access control**. As mentioned earlier, PFS users can enable various access control measures for their PFWs, e.g., a Ngrok user can enable access control to her PFW through HTTP Basic Auth, IP restrictions, or User Agent Filter. As detailed in §4.1, HTTP response from a PFW can help tell whether and which access control measure is enabled. Therefore, the PFW snapshots enable us to measure the extent of access control adoption, which turns out to be low. Specifically, only 3,252 (7.05%) reachable PFWs of Ngrok enable the HTTP Basic Auth, and only 55 (0.12%) reachable PFWs of Ngrok enable IP-based access control, , while none of reachable PFWs of Ngrok enbale the User Agent Filter. When it comes to Oray, it supports only IP-based access control. Once a PFW is enabled with IP-based access control, Oray will drop connections from unqalified IPs and make the PFW unreachable. Therefore, such kinds of PFWs are indistinguishable from truely inactive ones, which impedes us from measuring the adoption of access control for Oray PFWs.

**Usage**. The pDNS datasets allow us to learn the aggregate number of historical DNS queries for a PFW, which can be considered as a lower-bound approximate of a PFW's usage volume. Across the 6 million PFW domain names, a long-tailed pattern is observed for their distribution over the volume of DNS queries. Specifically, 87.23% Oray PFWs have DNS queries of fewer than 100, while it is 99.63% for Ngrok PFWs. Also, 2,569 Oray PFWs have more than 100K DNS queries while it is 55 for Ngrok PFWs. Note that some PFWs even have millions of DNS queries observed, e.g., *d43d-123-58-249-2.ngrok.io* with 2.6 million DNS queries, and *zolothx1999.imwork.net* with 154 millions of DNS queries. We further investigated PFWs with a high volume of DNS queries and found that most of them were redirected to error pages indicating that the PFWs were no longer available, while the left ones tended to be business websites of small companies. For example, a company called SINREY, which provides Internet audio broadcasting service, uses *sinrey.oicp.net* to make its internal broadcasting systems forwarded to the Internet. The cumulative distribution of PFWs per provider over the number of their historical DNS queries can be found in Appendix A.1.

**Origins**. Ideally, PFS will hide the origin of a web service while relaying its content to the public, which is what Oray does. However, for Ngrok, if the tunnel is provided as free, it will encode the egress IP address of the tunnelled internal service, as part of the PFW domain[5]. The domain of such a free tunnel will follow the pattern of *{random}-{origin-ip}.ngrok.io*, e.g., *f4e5-103-90-249-114.ngrok.io* has an IPv4 egress IP *103.90.249.114* and *1530-240e-404-8500-5284-14e1-41f0-73a3-985e.ngrok.io* with an IPv6 egress IP *240e:404:8500:5284:14e1:41f0:73a3:985e*.

Among 2,832,033 PFWs of Ngrok, 261,765 have their IP addresses encoded in their PFS FQDNs (fully qualified domain names), which result in 63,264 unique IPv4 addresses and 20,234 unique IPv6 addresses. We then queried ipinfo.io, a well-acknowledged IP intelligence dataset, to learn more about the origins of these PFWs. It turns out that these IPs are distributed across 173 countries and regions, and 5,414 different internet service providers (ISPs). The top 5 countries with the most PFWs are US (22.35%), CN (20.09%), DE(6.60%), RU (5.49%), and IN (4.53%).

We also found that one egress IP address may correspond to many PFWs. Specifically, A total of 104 egress IPs were identified as being assigned to 100 or more PFWs. For example, 98.160.245.127 was found to be the egress IP of 10,057 PFWs, while 211.21.127.126 was the egress IP of 2,824 PFWs. Further investigation shows that the egress IPs associated with many PFWs are registered under telecommunication services, suggesting that they may serve as dynamic gateway IP addresses shared by residential internet users and cellular device users.

**Categories**. Applying our PFW classifier (§3.3) to millions of PFW snapshots reveals that the snapshots of many PFWs are error pages provided by PFS providers. These error pages can be categorized as offline errors (e.g., when the PFS agent is offline), access control errors, and request/response errors (e.g., the resource is not found). Since these error pages are not the true web content of the respective PFW, they cannot be used to decide the true category of the respective PFW, and we therefore exclude them out before measuring the categories of PFWs. Among the 3,501,556 snapshots captured for 275,513 PFWs, 1,395,371 were predicted as error pages. After excluding these error-page snapshots, 2,106,185 are left and are considered as valid snapshots involving 149,763 PFWs.

---

5. https://ngrok.com/abuse

TABLE 5. The distribution of reachable PFWs across predefined categories.

| Category | Oray | Ngrok | Both |
|---|---|---|---|
| Industrial Control System (ICS) | 1.87% | 1.25% | 1.83% |
| IoT Controller and Devices (IoT) | 0.98% | 3.47% | 1.11% |
| Network Devices | 3.54% | 2.38% | 3.48% |
| Remote Desktop | 1.24% | 0.02% | 1.18% |
| Office Automation (OA) | 15.04% | 5.42% | 14.56% |
| Data Store | 2.35% | 4.69% | 2.46% |
| Code Repository | 0.62% | 4.00% | 0.79% |
| Network-Attached Storage (NAS) | 0.93% | 1.15% | 0.94% |
| Webserver Default Page | 5.83% | 8.38% | 5.96% |
| Blank Page | 68.37% | 13.70% | 65.61% |
| Others | 15.07% | 64.85% | 17.57% |

Given the PFWs with one or more valid snapshots (not error pages), 8,820 (7.60%) PFWs belong to categories that involve remote device access/control, namely, ICS, IoT, network devices, and remote desktop, while 28,673 (24.71%) reside in categories that may enable remote access to sensitive datasets, namely, office automation, data stores, code repositories, and network-attacked storage devices. Table 5 lists detailed statistics about the distribution of PFWs across their categories. For PFWs that may enable remote access to either critical datasets or devices, an in-depth analysis has been further conducted on sampled cases, which further highlights the security risks of exposing these sensitive PFWs to the public, and more details are presented in §5.1.

## 5. Security and Privacy Risks

### 5.1. The Exposure of Critical Websites

As PFWs of multiple categories (e.g., ICS and IoT) may enable remote access to either critical datasets or devices, we manually studied cases of these categories, with a focus on identifying their subcategories and understanding their access control mechanisms, i.e., whether and how a visitor is authenticated before being granted the access to data or devices. Given a sampled PFW of each category, we first look into its snapshots, which, in many cases, is sufficient for us to learn its subcategories and the enforced access control measures if any. Then, when the snapshots fail to provide sufficient information, we manually visit the PFW in a conservative manner so as to gain the necessary understanding while minimizing the visit footprint. Regarding access control, we are concerned with authentication (e.g., username/password authentication) and challenge-response tests (i.e., any types of Captcha). As a website can enforce challenge-response tests only after login failures and we are unable to conduct login attempts due to ethical considerations, our results on the adoption of challenge-response tests should be considered as a lower-bound estimate.

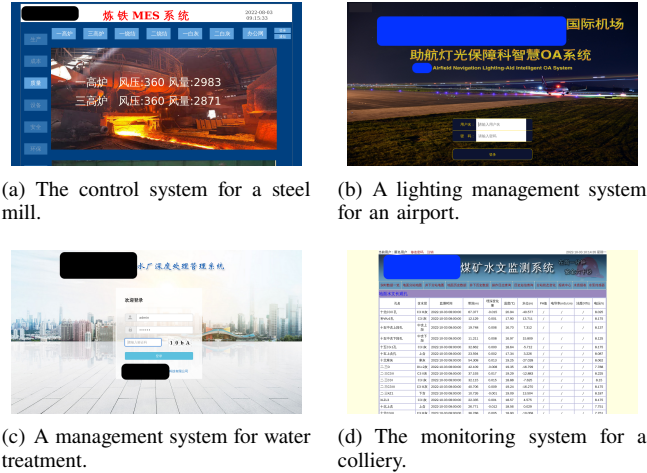

(a) The control system for a steel mill.

(b) A lighting management system for an airport.

(c) A management system for water treatment.

(d) The monitoring system for a colliery.

Figure 6. PFW cases in the category of ICS.

**Summary**. As summarized in Table 6, a non-negligible portion of PFWs fail to enforce any authentication mechanisms, therefore, leading to concerning risks in terms of data leakage or unauthorized device access. Below, we provide category-wise results in terms of subcategories and authentication meausures.

**Industrial control systems (ICS)**. In total, 2,056 Oray PFWs were classified as ICS, compared to 73 for Ngrok. Through manual labeling of a sampled set of 118 ICS cases, we have grouped them into 7 subcategories. The top subcategories include industrial or construction facilities (58.47%), transportation systems (12.71%), warehouses (10.17%), utilities facilities (7.63%), and mining facilities (5.08%). Figure 6 presents 4 typical cases that belong to different ICS subcategories, while the full list of ICS subcategories are listed in Appendix B.1.

Despite their critical roles, 13.56% ICS PFWs fail to enable any authentication mechanisms, which allows an unauthorized party to not only access sensor data, but also execute control commands. In addition, 77.97% support username/password authentication only, while 8.47% provide one or more alternative authentication factors in addition to username and password. We need to stress that the fraction of multi-factor authentication (e.g., 8.47% for ICS) should be considered as a lower-bound estimate, as a PFW may not activate more-factor authentication before the pass of the password authentication, for which, we are not allowed to test due to ethical considerations. Among ICS cases with authentication enforced, only 15.69% require some forms of challenge-response tests, suggesting that most ICS PFWs are vulnerable to password cracking attacks. One thing to recap, all access control results are learned through passively observing the PFW snapshots or visiting login pages of PFWs, which involve no login attempts.

**IoT controllers and devices (IoT)**. Following a similar process, we have further divided the IoT PFWs into a set of fine-grained categories (Appendix B.2), of which the top three are smart home controllers, video surveillance systems, and printers. Figure 7 illustrates these categories with concrete cases. Also, regarding authentication, 5.00% don't enforce any authentication, and only 3.51% require challenge-response tests for login attempts.

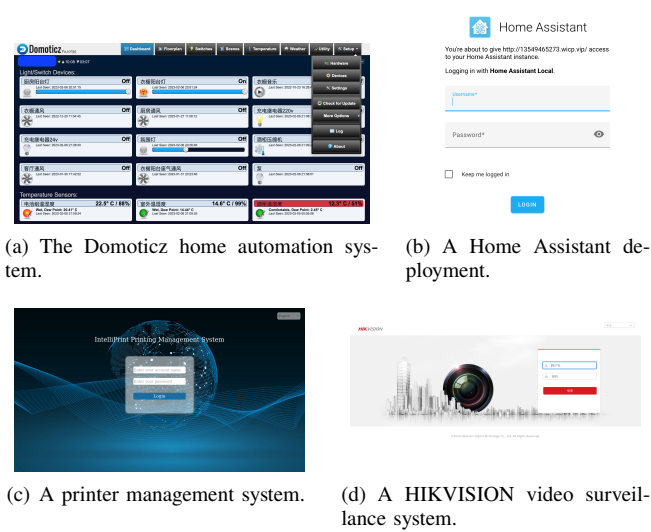

(a) The Domoticz home automation system.  (b) A Home Assistant deployment.

(c) A printer management system.  (d) A HIKVISION video surveillance system.

Figure 7. PFW cases in the category of IoT.

TABLE 6. THE STATS OF PFWS REGARDING THEIR AUTHENTICATION SCHEMES.

| Category | % No Auth[1] | % Only Passwd[2] | % CRAT[3] |
|---|---|---|---|
| ICS | 13.56% | 77.97% | 15.69% |
| IoT | 5.00% | 91.67% | 3.51% |
| Network Devices | 7.02% | 92.98% | 16.98% |
| Office Automation | 1.23% | 86.42% | 16.25% |
| Data Store | 65.15% | 34.85% | 0.00% |
| Code Repository | 0.00% | 100.00% | 0.00% |
| NAS | 3.57% | 96.43% | 0.00% |
| Remote Desktop | 37.50% | 62.50% | 0.00% |

[1] The fraction of PFWs with no authentication scheme enforced.
[2] The fraction of PFWs with only password authentication observed.
[3] The fraction of PFWs with challenge-response authentication tests.

**Data store**. Among PFWs of data store, 57.58% display a list of files and directories, and we consider these PFWs as directory index servers, 16.67% are self-hosted cloud storage systems which typically support not only file hosting and file transfer, but also file sync across devices and file sharing across users, 13.64% are collaborative document management systems, e.g., various wiki systems, 4.55% are traditional FTP-like file hosting systems, and 4.55% are monitoring and analytic dashboards. Also, among data store cases, 65.15% don't enforce any authentication, and none has challenge-response tests enabled during authentication.

**Code repositories**. 0.88% are code repositories, among which most are self-deployed version control solutions, e.g., GitLab and Gitblit. Following is Jupyter Notebook servers (18.09%), CI/CD systems (5.32%), and remote IDEs (3.19%), e.g., Visual Studio Code and RStudio. Among PFWs for code repositories, all have enforced authentication through username and password. However, none has required challenge-response tests for login attempts.

**Network devices and network-attached storage (NAS) devices**. Most PFWs of the category of network devices are found to be management consoles for routers from different vendors. Similar to code repositories, the majority have enabled username/password authentication, but in the meantime, fail to require challenge-response tests. Similar stats have also been observed for PFWs of NAS.

**Office automation (OA)**. 86.42% of OA support only username/password authentication, 12.35% provide one or more alternative authentication channels, and only 16.25% enable challenge-response tests for authentication.

**Remote desktop**. We found two typical Remote Desktop applications – Sunlogin[6] (87.04%) and noVNC[7] (11.11%). All Sunlogin PFWs require username and password for access, while for noVNC, 83.33% do not enforce any authentication. And none of either categories requires challenge-response tests when attempting to login.

### 5.2. Attacks to PFS Protocols

We have also identified a set of protocol vulnerabilities in Oray's PFS implementation, which incur non-negligible security risks to both PFWs and the local networks on which PFWs are hosted. Also, the identified security vulnerabilities have been responsibly disclosed to Oray along with acknowledgments received.

**Attacks against the data plane communication**. In Oray, the tunnel between the Oray data server and the Oray agent utilizes a customized application protocol over TCP. Further analysis revealed that it adapts the HTTP protocol with a message authentication code (MAC). And the MAC can be easily calculated without the need of knowing any secrets, as it only checks the length of the payload. This allows any intermediate hop between the Oray agent and the Oray server to perform a man-in-the-middle attack.

We have demonstrated this attack on our PFS testbed. In these experiments, we changed the firewall rules on the machine running the Oray agent, so that traffic between the Oray agent and the Oray server could be redirected through our mitmproxy node. And the mitmproxy worked in the transparent mode, so it could directly handle incoming packets at the network layer. On the mitmproxy node, we deployed scripts to modify the intercepted traffic and re-calculate the MAC. Through this process, our MITM node could arbitrarily modify the HTTP requests/responses being relayed through this tunnel, without notice from any involved parties. One thing to note is that it is not necessary for a real-world attacker to touch the Oray agent's firewall rules, as long as he controls one of the hops (e.g. a router) through which the PFW traffic will pass.

6. https://sunlogin.oray.com/en/embed/software.html
7. https://novnc.com/info.html

```
"phsl":"XX.oray.net:6061",
"mappings": [
    {
        "domain": "XX.xicp.fun",
        "punycode": "XX.xicp.fun",
        "servicehost": "127.0.0.1",
        "serviceport": 8001,
        "server": {
            "serverhost": "phfw-
                overseasvip.oray.net",
            "serverport": 6061,
            "feature": "tcp,udp",
            "serverudpport": 6061
        }
    }
]
```

Listing 1. Configurations from the Oray control server to the Oray agent.

**Attacks against the control plane communication**. We also identified and demonstrated another attack, which allows an attacker to use the PFS agent program as a stepping stone to expose internal infrastructure co-located with a PFW. Again, this attack exists only for Oray.

This attack surface exists in the control plane communication. When port forwarding rules are initially pulled from the control server, HTTPS is adopted. However, we have found that the Oray agent fails to perform proper server certificate verification, and an MITM attacker can successfully intercept this HTTPS connection and modify all the port forwarding rules that are passed from the control server to the agent. Specifically, Listing 1 shows the list of port forwarding rules that are used to tell the Oray agent regarding where and how to forward incoming traffic. All of these configurations can be modified by the MITM attacker. For example, by manipulating the *phsl* attribute, the attacker can modify the control server where the Oray agent to pull future configuration updates. Furthermore, through manipulating the *serverhost* and *serverport* fields, the attacker can instruct the agent to forward the incoming traffic to any other co-located internal web infrastructure, to any web servers under the control of the attackers, and even to any publicly available web services, as demonstrated by our control experiments. What is even worse, when pulling future configuration updates from the server specified in the *phsl* field, an unencrypted protocol is adopted, which allows any on-path hops to perform traffic manipulation without the need of a TLS certificate replacement.

The above vulnerabilities have been successfully demonstrated on our PFS testbed. In these experiments, we are able to instruct the Oray agent to forward the incoming requests to any IP or port, regardless of whether the IP is private or public. Also, as revealed by our experiments, the Oray agent will restart and pull the port forwarding rules again, as long as invalid response data is received from the Oray data server. This allows an MITM attacker to arbitrarily update the Oray agent with malicious port forwarding rules, at any time, by injecting invalid data into the data plane traffic. Regarding the attacking results, the most concerning one is that the PFS agent can be instructed to work as an internal stepping stone for the attacker to get access to any co-located internal web services. What's more, leveraging this attack, benign traffic can be transparently forwarded by the Oray agent to a malicious web server controlled by the attacker. Also, since traffic can even be maliciously forwarded to any public IP address, the Oray agent can become the scapegoat for denial-of-service attacks targeting a public web service. In this scenario, any visits towards the PFW will be forwarded by the misled Oray agent to the targeted web service. However, the magnitude of such DOS attacks is constrained by both the bandwidth allocated to the Oray agent as well as the volume of visiting traffic towards the PFW.

TABLE 7. THREAT STATS OF PFW DOMAINS AS LEARNED FROM VIRUSTOTAL.

| PFS | Queried | Analyzed | Ratio of Malicious PFW Domains [1] | | |
|---|---|---|---|---|---|
| | | | $\geq 1$[2] | $\geq 5$[2] | $\geq 10$[2] |
| Ngrok | 1.84M | 30K | 2.31% | 1.23% | 0.45% |
| Oray | 1.51M | 25K | 1.14% | 0.12% | 0.04% |
| Both | 3.34M | 55K | 1.78% | 0.73% | 0.26% |

[1] A PFW domain is considered as malicious as long as it has been alarmed by one or more detection engines underpinning VirusTotal.
[2] $\geq X$ denotes the ratio of malicious PFW domains that have been alarmed by $X$ or more detection engines, over all the analyzed PFWs.

Also regarding the attacking bar, the attacker doesn't have to be the ISP or even a nation-state actor. Instead, considering the Oray agent can be deployed in a movable device (e.g., a laptop, or even a mobile device), any public WIFI hotspots such as ones deployed in cafeterias, can serve as the attacking vantage points. Once malicious port forwarding rules are injected into the Oray agent and the device is subsequently introduced into a sensitive organization, co-located websites become accessible to attackers.

## 6. Abuse from Miscreants

Previous studies [8], [6] have revealed some separate incidents where PFSes have been abused by attackers, either to expose a compromised device to the public, or to hide their attack network infrastructures (e.g., the original servers of phishing websites). However, no studies have systematically profiled the abuse of PFS. In this section, we move one step further and provide a comprehensive analysis of the extent to which port forwarding services have been abused in various malicious activities. Given the 6.9 million PFW domains and their server IP addresses as observed from passive DNS, we extracted their threat reports from VirusTotal [8], a well-adopted threat intelligence platform, that provides threat reports for programs, URLs/domains, and IP addresses. However, due to the rate limit of VirusTotal, PFW domains and IPs were randomly sampled and queried at our

8. https://www.virustotal.com/

best effort. As the result, all PFW server IPs and a subset of over 3 million PFW domains have been queried.

**The maliciousness of PFW domains**. Table 7 presents the threat stats of PFW domains. And we can see that a non-negligible portion of PFWs have been alarmed as malicious by detection engines underpinning VirusTotal. Particularly, among Ngrok PFWs that have been analyzed by VirusTotal, 2.31% were reported as malicious while 1.23% have been further alarmed by 5 or more detection engines. For instance, *1ec6b9e8.ngrok.io* has been alarmed by 15 different detection engines for both phishing/fraud and malware distribution. Further investigations show that it was deployed to target Italian people with COVID-relevant frauds and victims would be lured to download a malicious Microsoft Excel file (e.g., *Corona virus documento protetto.xls*), which once opened, would infect the victim's machine with extra payloads (e.g., Revenge-RAT) [17], [18]. Actually, this is just one of over 14 different Ngrok PFWs abused by the same miscreants when conducting this fraud campaign [18]. One more example is *c9f44961.ngrok.io* which has received 17 alarms relevant to malware. Also, 5 programs were found to have ever contacted this domain name during execution and all of them were detected as malware.

Also, as elaborated below, we believe these ratio values of malicious PFWs are still lower-bound estimates. First of all, VirusTotal suffers from a low coverage of PFWs. As shown in Table 7, among all queried PFWs, only 1.64% Ngrok PFWs and 1.65% Oray PFWs have been analyzed by VirusTotal and thus have respective threat reports. Considering malicious PFWs can emerge and disappear very quickly, VirusTotal's low coverage of PFWs suggests it may have missed many malicious ones. One more factor resides in our observation that a single PFW domain name can serve malicious activities of different parties at different TCP ports, which however is only counted as a single malicious case when calculating aforementioned maliciousness ratios. Particularly, for Ngrok, a PFW domain name is meant to be used to tunnel raw TCP connections at different ports, if it matches a pattern such as *\*.tcp.ngrok.io* and *\*.tcp.eu.ngrok.io*. One example is *0.tcp.ngrok.io* which has received alarms for various malicious activities including malware distribution, phishing, and botnets. Also, over 18K different programs have ever communicated with this PFW domain name during their execution and most of them were detected as malware. Besides, the communication between these malware programs and this domain name spans the last 6 years (2017-2023), and such abuse is still ongoing, e.g., *5.tcp.eu.ngrok.io* was recently alarmed on November 22, 2023 for being contacted by a malware that has been alarmed by 63 out of 72 virus detection engines.

**The maliciousness of PFW IPs**. We then moved to profile the maliciousness of PFW server IPs. As PFW server IPs tend to be stable across time as well as being exclusively used for port forwarding, their maliciousness can help us better understand the abuse of PFS in malicious activities. As shown in Table 8, PFW IPs have been alarmed to a significant extent, e.g., among all analyzed Ngrok PFW IPs, 76% have been alarmed by one or more detection engines while over 55% have been alarmed by ten or more. Also, for both PFS providers, the malicious rate of PFW IPs is higher than that of the PFW domains, which is reasonable because a large volume of PFW domain names can be resolved to the same PFW IP, as a result of which, malicious traces of all these PFW domain names will be accumulated on this single IP address.

TABLE 8. THREAT STATS OF PFW SERVER IPS AS LEARNED FROM VIRUSTOTAL.

| PFS | Queried | Analyzed | Ratio of Malicious PFW IPs [1] | | |
| --- | --- | --- | --- | --- | --- |
| | | | $\geq 1^2$ | $\geq 5^2$ | $\geq 10^2$ |
| Ngrok | 189 | 142 | 76.06% | 69.01% | 55.63% |
| Oray | 28K | 28K | 2.88% | 0.51% | 0.02% |
| Both | 28K | 28K | 3.25% | 0.86% | 0.30% |

[1] A PFW server IP is considered malicious as long as it has been alarmed by one or more detection engines underpinning VirusTotal.
[2] $\geq X$ denotes the ratio of malicious PFW IPs that have been alarmed by $X$ or more detection engines, over all the analyzed PFWs.

Also, through manually studying the top cases that were alarmed most, we have confirmed that the maliciousness of these IPs should be exclusively attributed to PFS rather than any other co-located services or activities. For instance, *3.141.210.37*, a Ngrok PFW IP address, has been alarmed by 16 detection engines along with 2.4K communicating programs that are mostly malicious. Although it has ever resolved to a total of 11 domain names, the only malicious one is *6.tcp.ngrok.io*, a Ngrok PFW domain for tunneling raw TCP connections. Also, this IP was alerted by detection engines for serving as control & command servers (C2) for the njRAT malware [19], a well-adopted remote access trojan (RAT), that allows the attacker to remotely control and monitor a victim's computer via keystroke logging, accessing the camera, stealing in-browser credentials, etc. Apparently, Ngrok has been abused in this case by various njRAT operators to tunnel the communication between their real C2 servers and the victim machines. One specific indicator of compromise as captured by ThreatFox is *3.141.210.37:12336* [20]. Given the 2.4K communicating programs that have contacted this IP during execution, We manually looked into a randomly sampled subset and have thus confirmed that all of them are different variants of njRAT and all of them have contacted *6.tcp.ngrok.io* in the meantime.

One more issue that deserves some discussion is the difference between Ngrok and Oray in their PFW maliciousness as observed by VirusTotal. As shown in Table 7 and Table 8, Oray appears to have a lower ratio of malicious PFWs when compared with Ngrok, which can be attributed to at least two factors. On one hand, Oray is dedicated to customers in China while VirusTotal is known to have a low coverage for threat intelligence in China [9]. Besides, Oray requires real identity verification before using its service, which may likely have impeded many abuse attempts.

**Summary**. We can conclude with high confidence that port forwarding services are being abused to a concerning extent and in various malicious activities, particularly, tunneling communications for RAT programs, malware distribution, and phishing & fraud. Also, malicious PFWs can emerge and disappear very quickly while the underlying attacking servers are kept hidden, which renders many existing defensive mechanisms less effective, e.g., the reporting of a PFS IP:port as a C2 server IoC can soon become ineffective as the attacker can quickly migrate to a new tunnel.

## 7. Discussions

**Responsible disclosure**. We have responsibly disclosed our findings to relevant parties including the administrators of sensitive PFWs especially ones without access control enforced, as well as PFS providers. For vulnerable PFWs with contact information manually located, we notified their administrators through emails along with a simple questionnaire, which is designed to understand the causes of such PFW exposure, as well as administrators' attitudes towards the potential security risks. We have also communicated with Oray and disclosed the aforementioned protocol vulnerabilities, which have in turn been well acknowledged by Oray. More details can be found in Appendix C.

**Discovering PFW domain names**. As mentioned above, we query two seperate passive DNS databases to discover PFW domain names, which can counter the geographic bias that is inherent in passive DNS. In addition to passive DNS, certificate transparency (CT) logs provide an alternative way to discover PFW domain names that are exposed through HTTPS. To make it works, a prior condition is that PFS providers issue separate HTTPS certificates for different PFW domains rather than sharing wildcard certificates across PFWs, which is the case for Ngrok but not Oray. Besides, as some PFWs are exposed through only HTTP rather than HTTPS, they may be observed in passive DNS but not CT logs. Therefore, future works may combine both CT logs and passive DNS so as to achieve a higher PFW coverage.

**Preventing unauthorized port forwarding for internal websites**. As revealed above, many critical web services have been exposed to the public through PFS, and some of them even have no access control. Also, it is unclear whether such kinds of PFWs are exposed upon the authorization of their authentic administrators, since no mechanisms have been implemented by PFS providers to enable the authorization from the true administrators of PFWs. This allows the attacker with access to the internal network to expose sensitive internal web infrastructures that should only be available for internal visitors.

To address these issues, we propose a mitigation technique through requesting authorization from the authentic administrator before exposing an internal web service as PFW, which once deployed, can prevent a remote attacker from exposing local services even if it has gained priviledged access to the local network. This is achieved by utilizing protected confirmation dialog backed by trusted execution environments (TEE), which tend to be increasingly available in servers and consumer devices [21], [22]. The authorization process is triggered every time when the PFS agent is instructed to tunnel an internal web service. During the authorization, a confirmation dialog (i.e., a user consent dialog) is presented to the user interface of the device hosting the PFS agent. The confirmation (i.e., authorization) can only be given by either pressing a physical button of the device, or clicking the confirmation dialog protected by the TEE, which should prevent typical remote attackers as they have no physical access to the device. Then, no matter whether the authorization is granted or not, the authorization result along with the content in the confirmation dialog are securely signed by the TEE using its private key before being returned to the PFS agent, which in turn, sends the signed confirmation to the PFS server for verification. And the PFS server will refuse to expose a PFW until the signed confirmation is verified. When verifying the confirmation, the PFS server should first verify the signature is generated by a legitimate TEE hardware, then verify the confirmation dialog clearly states the port forwarding details, and lastly verify the authorization is granted. Such a signed confirmation can also be provided to the visitors of an exposed PFW for verification.

Assuming a remote attacker has compromised a local device, or even gain root permissions for the compromised device, as long as the TEE hardware is equipped, this mitigation solution can prevent such remote attackers from setting up PFS tunnels, as they don't have physical access to the compromised device. However, the TEE-based mechanism cannot prevent unauthorized but local attackers, e.g., a dishonest employee. As a local attacker may physically access a device, it can generate a valid and signed confirmation and thus circumvent such a defense. Also, this defense assumes the PFS providers are honest which we consider as reasonable. Also, we expect the modifications to both the PFS agent and the PFS server should be minor, as TEE APIs are available across platforms to facilitate seeking and verification of protected confirmation.

**Recommendations for administrators of internal web services**. As observed in our study, a large number of sensitive services in LAN are accessible on the Internet, exposing an unpredictable attack surface for the local network. The network administrators should be aware of this type of threat and implement a defence policy accordingly. Specifically, a threat model of zero trust is strongly recommended, and access to a critical web infrastructure, no matter whether it is from the inside or outside of the internal networks, should be strictly authenticated and authorized.

**New Apex domains.** We notice that Ngrok launched four new apex domains for its PFWs in April 2023 [23], which we believe doesn't invalidate any of our key results. Also, our PFS collector can be easily adapt to such updates by just updating the list of apex domains.

**Code and datasets release**. Considering many PFWs are sensitive or vulnerable, we decide not to publicly release the list of PFW domain names or their snapshots. Instead, we will delete the PFW snapshots once this study is finalized and only provide the PFW domain names and PFS apex domains under request and conditioned on strict background vetting. Also, we will publicly release the source code of our PFW collector and the resulting models of our PFW classifier.

## 8. Related Works

**Security studies on PFS**. Very few works study the security risks of PFS. As detailed in a security blog article [8], researchers from Huntress revealed how attackers had abused Ngrok to expose a compromised computer to the public. However, no details were provided regarding how such an abuse was captured and how prevalent such kinds of abuse are. Furthermore, when studying SMS spamming activities, Tang et al. [6] discovered the abuse of PFS for anonymizing the network servers of spamming activities. To the best of our knowledge, our study is the first work that has systematically vetted the security characteristics of the PFS ecosystem. For the first time, we have captured millions of PFW snapshots, demonstrated multiple protocol vulnerabilities, qualified and quantified the risks of PFWs that are of weak access control, and profiled various abuses from miscreants leveraging threat intelligence datasets.

**Security studies on network proxies**. A long line of works have studied the security of network proxies, especially virtual private networks (VPNs), the Tor network, open web proxies, and residential proxies. Regarding VPNs, Ikram et al [24] evaluated the security characteristics of 283 Android VPN apps along with concerning security issues discovered, e.g., the misuse of insecure tunnelling technologies and TLS traffic interception. Similarly, Taha et al. [25] conducted a security auditing for 62 VPN providers and discovered that many VPN providers made false claims regarding the geographic diversity of their VPN servers. Given the issues of existing VPN protocols and deployments, novel VPN protocols have been proposed, particularly, Wireguard [26].

In addition to VPNs, the Tor network [27] has also attracted much attention. Some studies focus on evaluating the robustness of the Tor network through novel attacks [28], [29], [30], [31]. Particularly, Murdoch et al. [28] proposed a novel traffic analysis technique to effectively infer whether a given relay is used to relay an anonymous traffic stream. Besides, Overlier et al. [29] demonstrate the feasibility of inferring the location of a hidden onion service leveraging a single malicious relay. Another set of works [32], [33] explore how to conduct denial of service attacks against the Tor relays, e.g., the Sniper attack [32]. As the Tor network getting increasingly adopted, many studies moved to profile the Tor network in terms of the general usage as well abuse from miscreants [34], [35], [36], [37], [38]. Given the abuse of the Tor network in malicious activities, traffic relayed by the Tor network suffers from service discrimination or blocking [39] by popular online service providers. To bypass Tor exit blocking, Zhang et al. [40] proposed the use of short-lived proxies as alternative egress points for Tor traffic and they name such proxies as *exit bridges*.

Furthermore, multiple security risks have been identified for open web proxies [41], [42]. Particularly, Tsirantonakis et al. [41] discovered that 5.15% of the evaluated open HTTP proxies were found to have performed content modification or insertion. Another kind of network proxy is residential proxies, and multiple studies [43], [44], [9] have revealed the suspicious recruitment of residential proxies [44] as well as the abuse of residential proxies in malicious activities [43], e.g., advertisement fraud. Particularly, Mi et al. [44] studied how mobile devices got recruited as residential proxies. Also, Chiapponi et al. [45] explored how to detect traffic flows relayed by residential proxies from the side of traffic destinations.

**Webpage classification**. Webpage classification techniques have been comprehensively explored in previous studies from multiple aspects. First of all, different classification algorithms have been applied to webpage classification, e.g., k-nearest neighbors [46], support vector machine [47], [48], [49], genetic algorithms [50], one-class classification algorithms [51], ensemble learning [52], and different neural network architectures [53], [54], [55], [14]. Also, multiple feature engineering techniques have been proposed and evaluated. Particularly, Several works [56] focused on extracting features from the URL of a given webpage, while Shen, et al. [57] proposed extracting features from the webpage summarization. Furthermore, webpage classification techniques have been applied to the security domain, e.g., the detection of malicious URLs [15], sensitive webpage classification [16], and phishing webpage classification [58], [59]. Moving forward, in this study, we have applied webpage classification techniques to decide which security-sensitive category a PFW belongs to, along with a good classification performance achieved.

## 9. Concluding Remarks

In this study, we have conducted the first of its kind security study on the ecosystem of port forwarding services (PFS). This is made possible through designing and implementing a novel methodology, to automatically discover and snapshot port-forwarded websites (PFWs) at scale, classify PFWs into pre-defined categories by considering multi-modal webpage elements, as well as identifying and experimenting novel attack scenarios. As the result, multiple inspiring security findings have been distilled. Particularly, many critical websites have been port-forwarded to the public with either weak or no access control; the port forwarding protocol of a major PFS provider has multiple vulnerabilities identified and demonstrated, and PFSes have been extensively abused in various malicious activities. To conclude, non-negligible security and privacy risks have been introduced by the PFS ecosystem, to address which, more research and engineering efforts should be invested.

# Appendix A.
# Complimentary Measurement Results

## A.1. The Usage of PFWs

Figure 8 presents the cumulative distribution of PFWs over the DNS queries.

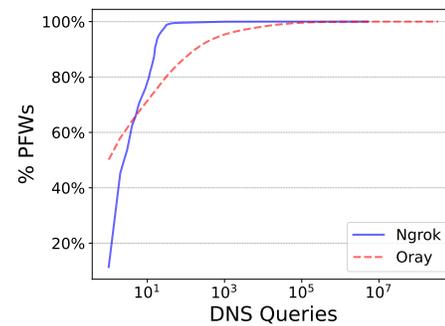

Figure 8. The cumulative distribution of PFWs over their DNS queries.

# Appendix B.
# The Security Risks of PFWs

## B.1. The ICS Subcategories

Table 9 presents the full list of 7 subcategories for the PFW category of industrial control system (ICS).

TABLE 9. THE SUBCATEGORIES OF PFWS THAT BELONG TO INDUSTRIAL CONTROL SYSTEM (ICS).

| Subcategory | Ratio |
|---|---|
| Industrial Manufacturing | 58.47% |
| Traffic | 12.71% |
| Warehouse | 10.17% |
| Livelihood | 7.63% |
| Mining | 5.08% |
| Logistics | 3.39% |
| Others | 2.54% |

TABLE 10. THE SUBCATEGORIES OF PFWS THAT BELONG TO IOT CONTROLLERS AND DEVICES (IOT).

| Subcategory | Ratio |
| --- | --- |
| Home Automation System | 63.33% |
| Video Surveillance System | 30.00% |
| Printer Management System | 6.67% |

## B.2. The IoT Subcategories

Table 10 presents the full list of 3 subcategories for the PFW category of IoT controllers and devices (IoT).

# Appendix C.
# Responsible Disclosure

## C.1. To Administrators of Sensitive PFWs

As detailed in §5.1, 32.31% PFWs are classified into security sensitive categories, e.g., industrial control system. It makes us wonder whether the administrator of such a PFW is aware of the exposure of the website. If not, for what reasons the website got exposed through PFS? If yes, to what extent do the administrator understand the potential security risks? To answer these questions, and to responsively disclose our research results to relevant parties, we collected contact information (email addresses) for sensitive PFWs at our best efforts, designed and implemented an online questionnaire, notified the respective parties of the exposure of their internal websites, and invited them to participate in our questionnaire. More will be elaborated below.

**Collecting email addresses of sensitive PFWs**. The PFW organization tends to have contact information in the form of either a telephone number or an email address, or even both. When collecting contact information, we consider only email addresses since we can easily automate the email communication. Across PFWs, the general procedure starts from visiting a PFW and identifying the organization behind it. Sometimes, the contact information is also available on the website along with the organization name. But more frequently, we have to search the Internet with the organization name to further locate the contact information. What is even worse, there is even no organization name for many PFWs. At our best manual efforts, one or more email addresses have been identified for 119 PFWs.

**The questionnaire**. Our questionnaire is designed in a concise manner so that the participant can complete it within one minute. It includes only 9 questions, aiming to gather information for three research questions: 1) What led to the exposure of the PFW? 2) To what extent do the participant understand the security risks? 3) What actions do they plan to take given our disclosure? Also, we provide the questionnaire as well as the email in either English or Chinese, depending on the language of the respective PFW. the English version of this questionnaire is given in Appendix C.3.

**The disclosure process**. During the disclosure, emails were composed leveraging the email template presented in Listing 2. In each email, a unique link is embedded for the receiver to participate in our online questionnaire. In addition, for each PFW email address, the same disclosure email will be sent up to three times in case the emails were missed. There is also a one-week interval between sending duplicate emails to the same email address, to avoid potential spamming.

**The results**. By this writing, we have completed the disclosure process. However, among the 230 PFW email addresses we have contacted, only 1 replied without any concrete feedback. Also, 68 opened our online questionnaire, but none of them participated in the questionnaire yet.

## C.2. To PFS Provider

Regarding the MITM vulnerabilities discovered in Oray's port forwarding protocols, we have contacted Oray for responsible disclosure and they have confirmed these vulnerabilities along with a bug bounty rewareded to us.

```
We are cybersecurity researchers from
    XXX. In our recent security research
    on port forwarding service {PFS}, we
    found that a security-sensitive
    website of your organization was
    exposed to the public network through
     {PFS}.  In accordance with the Code
    of Ethics for Security Research, we
    sincerely inform you of this exposure
    . Here is more specific information:
    your website was exposed through the
    port-forwarding domain of {PFW domain
     name}, and the screenshot of the
    landing page of the exposed website
    is attached as below.

Feel free to reply to this email if you
    have any questions. We would
    appreciate it very much if you can
    participate in our 1-minute survey
(a simple questionnaire).

Looking forward to hearing from you!
```
Listing 2. The email template to contact the administrator of a PFW.

## C.3. The Questionnaire Designed for Responsible Disclosure

1) Information of the subject.
   a) Your institution name.
   b) Your contact phone number.
   c) Your contact email.
2) Does the website disclosed in the email belong to your company?

a) Yes.
   b) No.
3) Did your organization proactively expose the website to the public network?
   a) Yes.
   b) No.
4) If not, was the exposure of the website a misoperation by your employees?
   a) Yes.
   b) No.
5) If not, was the exposure of the website a result of some cyberattacks?
   a) Yes.
   b) No.
6) If your company voluntarily exposed the website to the public network, what was the exposure intended for?
   a) Serve public network users.
   b) Serve remote employees.
   c) Both.
   d) Neither.
7) To what extent do you think that exposing the business website to the public network environment will pose a security risk?
   a) Yes, and there are great security risks.
   b) Yes, but the security risks are acceptable.
   c) Yes, but the security risks are low or non-negligible.
   d) No, there is no security risk.
8) Do you plan to stop exposing this website to the public network?
   a) It has already been stopped.
   b) Plan to stop it.
   c) Not going to stop it.
9) Your suggestions or other feedback.[9]

---

9. This question is designed as an open-ended question.